# Enriching the scholarly metadata commons with citation metadata and spatio-temporal metadata to support responsible research assessment and research discovery

**Daniel Nüst[1*], Gazi Yücel[2], Anette Cordts[2], Christian Hauschke[2]**

[1]Institute for Geoinformatics, University of Münster, Germany

[2]TIB – Leibniz Information Centre for Science and Technology, Hannover, Germany

* Correspondence: Daniel Nüst, <daniel.nuest@uni-muenster.de>

**Abstract**

In this article, we focus on the importance of open research information as the foundation for transparent and responsible research assessment and discovery of research outputs. We introduce work in which we support the open research information commons by enabling, in particular, independent and small Open Access journals to provide metadata to several open data hubs (Open Citations, Wikidata, Open Research Knowledge Graph). In this context, we present The OPTIMETA Way, a means to integrate metadata collection, enrichment, and distribution in an effective and quality-ensured way that enables uptake even amongst small scholar-led publication venues. We have designed an implementation strategy for this approach in the form of two plugins for the most widely used journal publishing software, Open Journal Systems (OJS). These plugins collect, enrich, and automatically deliver citation metadata and spatio-temporal metadata for articles. Our contribution to research assessment and discovery with linked open bibliographic data is threefold. First, we enlarge the open research information data pool by advocating for the collection of enriched, user-validated metadata at the time of publication through open APIs. Second, we integrate data platforms and journals currently not included in the standard scientometric practices because of their language or lack of support from big publishing houses. Third, we allow new use cases based on location and temporal metadata that go beyond commonly used discovery features, specifically, the assessment of research activities using spatial coverage and new transdisciplinary connections between research outputs.

**Enriching the scholarly metadata commons**

1      Introduction

Open research information (ORI) provides a foundation for transparent and responsible research assessment and effective discovery of research outputs. Therefore, the quality, openness, extent, and scope of ORI should be as high as possible. However, especially for small and independent Open Access journals, it is difficult to collect publication metadata and deposit it in open research information data hubs. This is especially the case for more complex metadata that goes beyond the simple properties of individual records, such as title or publication date, by making connections between scientific publications. In this work, we introduce a concept for eliciting the rapid publication of verified open publication metadata, at the most effective and efficient moment in the publication process. We call this concept The OPTIMETA Way and implement it with two plugins for the publishing software Open Journal Systems (OJS) developed by the Public Knowledge Project (PKP; https://pkp.sfu.ca/). These plugins cover two distinct types of metadata that have very high potential to support more responsible research assessment and more powerful research discovery, but that are not yet widely available: verified and complete open citation metadata and spatio-temporal metadata.

The main contribution of this work is a concept for effective publication metadata collection and publication that (a) balances data quality with the need for manual user interaction, (b) targets specific points in the publication process when the motivation and expertise of the stakeholders are provided, and (c) enables independent journals to create innovative metadata that goes beyond the common standard of even large commercial publishers. This concept is here demonstrated through prototypical implementations, the OPTIMETA services. The OPTIMETA services target two areas of publication metadata that unleash the integrative power of spatial and temporal relations between research outputs and facilitate the much-needed availability of open citation information to tackle the overwhelming amount of scientific literature.

This paper is structured as follows. First, we briefly introduce the foundational background work considering, in particular, the potential diversity of the audience. Then, we present a concept and implementation strategy for innovative publication metadata. To conclude, we relate our ideas and products to the scholarly metadata commons, research assessment, and research discovery concepts before discussing the benefits, limitations, and future directions of the work.

2      Related Work

2.1      Open bibliographic metadata as a basis for responsible research assessment

Research is resource-intensive, and the actors organising and funding it have an interest in screening and evaluating the research results generated from these efforts. Thus, research assessment is being conducted for various reasons, among them, because of a "lack of trust" in academia and a desire to facilitate improved resource allocation. Such assessments are often presented as a metadata-driven quantification of the research process based on elaborate frameworks and assessment rules. They are usually based on output and reception-based indicators, which in turn can only be produced if suitable metadata is available for this purpose.

**Enriching the scholarly metadata commons**

In recent years, the traditional methods of assessing research have been fundamentally challenged from various sides. The role of bibliographic metadata is one important issue that has been raised, for example, by two widely discussed initiatives, as such material has to be made available through open licences in order to enable fair and responsible research evaluation. Both initiatives comment on the data basis required for responsible assessments. In 2012, several influential players in scholarly publishing met at the Annual Meeting of The American Society for Cell Biology in San Francisco, CA, USA (The American Society for Cell Biology 2012) to discuss emerging issues related to research evaluation. The result of this meeting was the San Francisco Declaration of Research Assessment (DORA, see Cagan 2013). By July 2022, almost 22,000 individuals and organisations in 158 countries had signed the declaration. DORA gives clear instructions for publishers and suppliers of metrics to be "open and transparent". One key action recommended by DORA is that publishers should "remove all reuse limitations on reference lists in research articles and make them available under the Creative Commons Public Domain Dedication licence" (recommendation 9), while metrics suppliers should make the data and methods underlying their metrics available under an open licence (recommendation 11 and 12). In the second initiative, launched in 2015, the Leiden Manifesto (Hicks et al. 2015) took this further with more detailed instructions on how and, in particular, why the metadata used for assessment should be shared. Two of the resulting principles, 4 ("Keep data collection and analytical processes open, transparent and simple") and 5 ("Allow those evaluated to verify data and analysis"), are particularly relevant in the context of the present article.

In order to comply with the principles and recommendations for responsible research assessment, data sources must meet various criteria. The research information - that is, metadata about actors, events, processes, and output related to research activities - must be findable and accessible if the publications are to be evaluated, and the records must contain the metadata in a form that allows legal and technically easy reuse. These criteria (findability, accessibility, interoperability, and reusability) are described by Wilkinson et al. (2016) as the so-called "FAIR Guiding Principles for scientific data management and stewardship". Taken to its logical conclusion, this means that any research assessment should be based exclusively on metadata that is derived from publicly available data sources, is openly licenced, and is published using open standards.

## 2.2 Open bibliographic metadata and spatio-temporal metadata as part of open research information

Open Research Information (ORI) is an emerging term used to describe metadata that complies with the above criteria for data sources intended for use in responsible research assessment. Recently, various actions have been taken and initiatives launched with the aim of more precisely defining and promoting ORI. For example, in 2018, the 14th International Conference on Current Research Information Systems (euroCRIS 2018) focused on the "FAIRness of Research Information". In 2020, a German-Ukrainian project discussing the topic of "FAIR Research Information in Open Infrastructures" with international experts (Kaliuzhna and Altemeier 2021) led to the development of high-level criteria that applied the FAIR principles to research information (Hauschke et al. 2021a). Bijsterbosch et al. (2022) described the "Seven Guiding Principles for Open Research Information" and provided a more detailed analysis of "Trusted and transparent provenance", "Openness of

**Enriching the scholarly metadata commons**

Metadata", "Openness of Algorithms", "Enduring access and availability", "Open Standards & Interoperability", "Open collaboration with Third parties", and "Academic sovereignty through governance".

Open bibliographic metadata is an important part of ORI, especially in relation to output-oriented research assessment. From a broader perspective, this relates to any metadata that describes published works, e.g., journal articles, conference proceedings, monographs, or edited books. Being the foundation and fuel of the publishing and library worlds, bibliographic metadata is produced by authors, editors, librarians, and many others involved in the dissemination of scholarly output. The conventional bibliographic metadata types are, e.g., reference lists, abstracts, author affiliations, author identifiers, and licences. Besides research assessment, bibliographic metadata is used in several other ways such as the creation of scholarly knowledge graphs (e.g., Jaradeh et al. 2019; Manghi et al. 2021; Priem et al. 2022; Turki et al. 2021) and in library catalogues and bibliographic discovery services (Gonzales 2014).

Recently, further types of bibliographic metadata, spatial and temporal metadata, have gained attention (Niers and Nüst 2020). Spatial and temporal metadata can deliver precise information about the location and time period that is covered in a publication. These metadata enable connections to be drawn between different research outputs, however, the availability and use of spatio-temporal metadata in ORI are currently sparse. The integrative potential of time and space is underutilised for publications (Niers and Nüst 2020) as well as for data (Garzón and Nüst 2021a) and current research information systems (CRIS) platforms.

Several initiatives and projects are working on enriching the scholarly metadata commons. While an exhaustive review of all activities can not be given here, some key examples include Rasberry et al. (2019), who show how Wikidata can be used as a source for scholarly metadata through its frontend Scholia. They even discuss location data, though primarily in relation to the author and their institutional address as coordinates. Nielsen et al. (2018) expand on the use cases for discovery based on location information and also describe the opportunities for querying using locations mentioned as topics in articles in Scholia using point and polygon features from Wikidata. Lauscher et al. (2018) argue that libraries should play an important role in the production and curation of scholarly metadata and prove the feasibility and effectiveness of the strategy for the case of citation metadata from printed books in the social sciences. The Ukrainian Open Citation Index is an example of the nationwide collection of citation metadata from academic publishers (Nazarovets 2019). On an international level two initiatives have gained a lot of traction: The Initiative for Open Citations (https://i4oc.org/) and the Initiative for Open Abstracts (https://i4oa.org/). Nevertheless, for all the merits of these projects and initiatives, there is still much to be done, especially for small, independent and scholar-led journals. For these journals, the citation metadata and spatio-temporal metadata have great potential because given these metadata are available in a structured and machine-readable format and are accessible in open bibliographic databases, they enable connections to be drawn between different publications and, thus, improve the visibility of the contributions made by the large number of scholar-led Open Access journals.

## 2.3 Citation metadata

Citation metadata is metadata that expresses how one document refers to another. The idea of recording cross-references between documents was raised as early as 1952 by Eugene Garfield in a talk to the Maryland Section of the American Chemical Society. He stated "If authors would provide the CA abstract number with each bibliographical citation, I can assure you that CA abstracts in the future would be much more informative by providing cross-references to related abstracts" (Garfield 1952). Later, he went on to create the Science Citation Index, which evolved into the citation database Web of Science and inspired many similar products in the decades that followed.

The description of citation metadata is a simple relation between two objects and several standards and definitions for different applications have been developed over time. To illustrate the diversity of the various efforts, we discuss three examples involving different types of citation metadata. Starr and Gastl (2011) present the initial way in which DataCite depicts relations between publications and research datasets. The basic assumption is that every entity, every research output in the DataCite metadata schema is described using a minimum number of mandatory fields: Creator, Title, Publisher, Identifier and Publication Year. Relationships between individual entities can then be constructed in various ways. IsCitedBy (and its inverse property Cites) tracks relationships between works that cite each other. This approach is content-agnostic. Peroni and Shotton (2012) developed the Citation Typing Ontology (CiTO), in which the citation still connects two works, but the citation itself is considered an entity that can be described by various properties independently. This allows for a much more detailed description of various types of relations and the representation of characteristics such as in-text citation frequency.

Over time, the standards for describing citation relationships for specific types of works have also emerged. A recent example is the citation of research software, for which Smith et al. (2016) have developed principles that seek to capture the specifics of this type of work. For example, it must be possible to address the versioning common in software development, to describe specifically whether a particular version of a software is cited, any of its versions, or its latest version.

## 2.4 Spatio-temporal metadata

Geospatial metadata is metadata for geographic data and information (Wikipedia 2022). There is a wide variety of standards, formats, and tools, ranging from public and industry standards for encoding geospatial metadata, such as the complex ISO 191** suite of standards, to catalogues collecting and serving geospatial metadata online (Federal Geographic Data Committee). While these types of data are often relevant, research output and publishing datasets are becoming more common and more widely acknowledged. Our study focuses on the geospatial properties of more classical research outputs: research papers. Therefore, we use the term *spatio-temporal metadata* when referring to the metadata of a spatial or temporal nature that describes textual and graphical research outputs. This type of metadata can refer, for example, to the spatial extent or the so-called area of interest that a scientific article investigates, or to the time period for which data were analysed. This approach has been demonstrated previously by JournalMap (Karl et al. 2013), albeit

**Enriching the scholarly metadata commons**

with considerable limitations (Hauschke et al. 2021b). Furthermore, the most commonly used research data repositories (e.g., Zenodo, OSF, and Figshare) do not explicitly support spatial metadata. Only a few research data repositories are tailored to handle georeferenced data, such as Pangaea (https://www.pangaea.de/about/) and CKAN with its spatial extension (https://ckan.org/features/geospatial/). Dataverse only supports vector data in the outdated format Shapefile (https://guides.dataverse.org/en/latest/developers/geospatial.html?highlight=geospatial). This lack of support is possibly due to the fact that the handling of geospatial data was not a common feature of database software (except with the additional software extensions) or the focus of expert specialisation.

Metadata of such a kind is relevant for a broad variety of scientific fields. In the natural and life sciences, observation data, model data, habitats or the sites of finds represent the most obvious points of connection (concerns, e.g., Earth science, geology, oceanography, meteorology, ecology, zoology, and botany). For the social and applied sciences, the connections to humans and their physical areas of activity are relevant in, for example, the medical and health sciences, agricultural science, economics, engineering, sociology, political science, and more. The theoretical work in the formal sciences (e.g., mathematics, logic, computer science) or the small-scale and theoretical physical sciences (physics, chemistry) are, as expected, less interested in geospatial metadata, although interdisciplinary work or research that features some element of application often has a real-world connection, i.e., a location in space and time.

Turning our attention back to platforms for handling research publications, neither big commercial solutions for CRIS, such as Pure (Elsevier) or Converis (Clarivate Analytics), nor widely used open projects, such as DSpace (Smith et al. 2003), support spatio-temporal metadata for any record type. VIVO (Conlon et al. 2019) has a property for (vivo:geographicFocus) describing the spatial component of a given research activity or output, but it is not widely used. Developer documentation and code repositories show interest in the topic, for example, DSpace lists several occurrences where a spatial search was suggested or prototyped (cf. https://github.com/DSpace/DSpace/pull/511). The same lack of support occurs in relation to discovery platforms (e.g., ScienceDirect, Google Scholar) and publishers' websites. The only explicit modelling of spatio-temporal information in the publishing domain is the spatio-temporal fields in the Dublin Core specification *DCMI Metadata Terms*, **coverage** being the most important among them (https://www.dublincore.org/specifications/dublin-core/dcmi-terms/terms/coverage/). However, this field is not particularly useful for machine-readable information exchange, as it may hold any type of spatial or temporal metadata, be it prose, coordinate pairs, or textual encoding of complex geometries or time periods. DCMI Metadata Terms use some alternative terms, e.g., jurisdiction or location, but these do not seem to be implemented in the platforms mentioned here. Thus, all these platforms include inexplicit or not directly usable spatial information, e.g., in the form of addresses for researchers, location names in paper abstracts, excavation site coordinates in the full text, or remote sensing imagery as figures in the supplementary material. Temporal metadata is more common, not least because it is simpler in nature and readily supported by any database management system. However, the temporal fields are often focused on publication metadata (when was the resource created or published?) rather than on the content. Similarly, this kind of temporal information is hidden in titles, abstracts, or full texts.

**Enriching the scholarly metadata commons**

This completely neglects the potential for spatial and temporal information to act as an *integrator*, e.g., as Kuhn (2012) argued in relation to transdisciplinary research. The potential has been pointed out in the past, both in relation to creating new connections between publications (Niers and Nüst 2020) and in relation to data (Garzón and Nüst 2021a; Garzón and Nüst 2021b). In general, spatio-temporal metadata currently does not play an important role in ORI.

## 3 The OPTIMETA Way

### 3.1 Approach

The introduction presented the challenges and benefits involved in enriching the scholarly metadata commons and how this connects with the responsible assessment and effective discovery of research. In order to tackle some of the challenges of capturing and disseminating high-quality and useful metadata for research outputs, we developed an approach to strengthen the Open Access publishing system through open citations and spatio-temporal metadata - *The OPTIMETA Way*. This concept recognises the conflict involved in metadata creation and usage during the publication phases of the research cycle.

Firstly, the benefits of creating high-quality metadata are intangible for authors, although they are the most knowledgeable source for most of the relevant information. However, this may change if the reasons for providing the metadata are communicated clearly, such as enabling a more responsible assessment of their work and better connection with other disciplines for evaluation and discovery purposes. Nevertheless, the way the metadata are captured should be intuitive and engaging, keeping in mind James Frew's laws: *"Frew's first law: scientists don't write metadata. Frew's second law: any scientist can be forced to write bad metadata."* (Hey 2015). Secondly, the large metadata owners are currently the big scholarly publishers. They have a strong interest in building their business and making a profit based on such data (Pooley 2022; Brembs 2021; Franceschi-Bicchierai 2022) and little interest in contributing everything they can to the creation of knowledge or addressing the need for technical innovation. Instead, innovation must be pursued by those who have an interest in having an open and free scholarly metadata commons, such as university publishers or independent journals, even though they have limited resources.

In acknowledgement of these conflicts and challenges, The OPTIMETA Way is output-oriented and focuses on enabling the essential, if currently relatively powerless, stakeholders in academic publishing to capture and distribute scholarly publication metadata themselves. The approach is **"OPTImal"** in the sense that it begins with small improvements with the potential to generate the biggest benefits: the potential impact of high-quality citation metadata on research assessment and quality-ensured metascience is huge. Furthermore, the novelty of spatio-temporal metadata and its integrative potential make it attractive even at a rudimentary level, for example, when simple geometries representing articles are shown on the same map for visual inspection and discovery.

**Enriching the scholarly metadata commons**

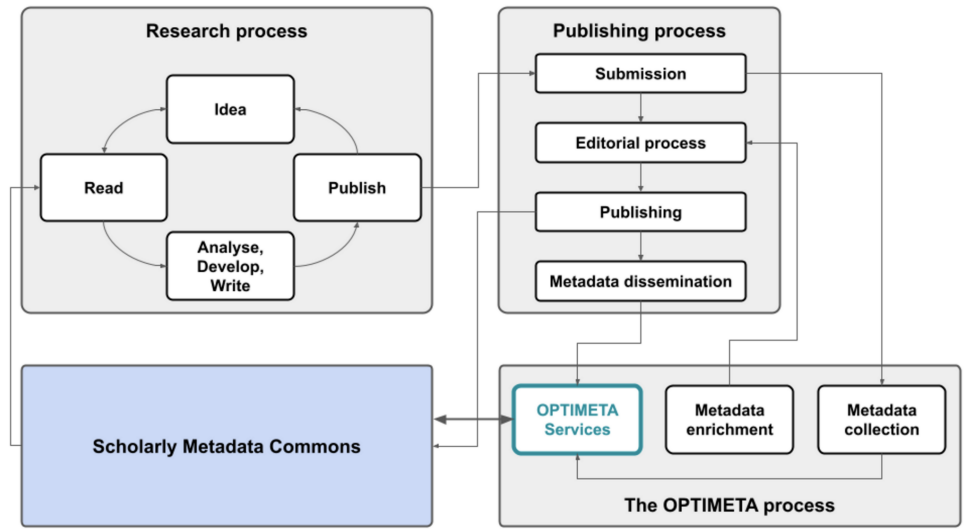

**Fig. 1.** Stages of the publication process, the generic research process, and The OPTIMETA Way with their connections.

This approach is **output-oriented** as it will assist metadata creators through automation and intuitive user interfaces and, thereby, avoid requiring additional strenuous, time-consuming, or unwelcome tasks. By tapping into open data sources, the software enriches the user input which is then provided for inspection. The approach further targets the quickest gains compared to the required effort and thus does not have to be comprehensive. The metadata will be created at a point in the publication process when authors are most willing to fulfil all administrative requirements, that is, while submitting an article for review and then, eventually, publication. This is also the last point in time at which engaged professionals (reviewers, editors, publishing staff) examine the metadata critically and at which the publication of output, including metadata, is already a core part of the process. Furthermore, the output is always checked by humans during this process and, as the data are not being produced by an algorithm, higher-quality metadata can be returned to the data sources used for enrichment.

Finally, the **facilitation** happens through The OPTIMETA Way. This approach enables independent journals and small publishers, who often work with open-source software platforms, to actively engage in structured metadata collection and distribution without expert knowledge. A crucial step in enabling this is the deposition of metadata in open data hubs, which will allow journals with the shared mission of disseminating knowledge to contribute to the bigger goal of an open scholarly publication metadata commons. Thus, the impact of The OPTIMETA Way will be bigger than the sum of its parts.

Fig. 1 is a schematic representation of The OPTIMETA Way. One side shows the simplified research cycle including the research activities and their translation into written output. The other side shows a journal's publication process from submission to editing to publication and the associated or subsequent dissemination of metadata. The OPTIMETA Way means that we support the generation, enrichment, and dissemination of metadata during the stages of the journal publication process in which the metadata is already a focus. In this way, this approach avoids any retrospective editing and post-publication tasks for researchers.

## 3.2 Implementation examples

### 3.2.1 Overview

Open Journal Systems (OJS) is the most widely used journal publishing software in the world. It was developed to make scientific publishing easier and more effective (Willinsky 2005). It organises the complete publishing workflow from submission to review, from proofreading to production. The software is open source and is continuously being improved by a global community. It is currently available in version 3.3. OJS provides the core functionalities discussed above, including metadata creation, editing and export. Additional functionality can be added to OJS through plugins that may be installed manually or using the so-called Plugin Gallery (https://docs.pkp.sfu.ca/plugin-inventory/en/). These plugins can extend or alter any step in the OJS submission and publication workflows and add new features to the editorial backend or the system's front end.

We have realised The OPTIMETA Way through two plugins: the OPTIMETA citation plugin and the OPTIMETA geoplugin ("citation plugin" or "geoplugin" for short). Together, these two plugins implement the **OPTIMETA Services** described in Fig. 1. Both plugins collect metadata during the submission workflow and enhance it with data from open scholarly data sources. They then present the person submitting the data with the enhanced information before, ultimately, exposing the information on both the OJS website and the external data sinks. The connection to external data sinks can be made in near real-time, in the sense that information is deposited actively in connection with events in the publishing workflow. Alternatively, other platforms can be used as a harvesting mechanism to regularly retrieve the published metadata from OJS. Fig. 2 shows the data sources and sinks that are currently supported by the plugins, as well as those that may be supported in the future. The data sources on the left are marked in red. Both plugins rely on user-contributed metadata that is enriched with external references and additional data. The targeted metadata platforms and aggregators, or data sinks, are shown on the right and marked in green. The journals and university publishers collaborating with the project are listed at the bottom (see the full list of names and links at https://projects.tib.eu/optimeta/en/).

The plugins are available as public beta releases and can be freely downloaded and installed from our public GitHub repositories at https://github.com/TIBHannover/optimetaCitations/ and https://github.com/TIBHannover/optimetaGeo. The plugins will be improved based on feedback from our OPTIMETA project partners and the OJS community (users and developers) that is focused, in particular, on the user experience.

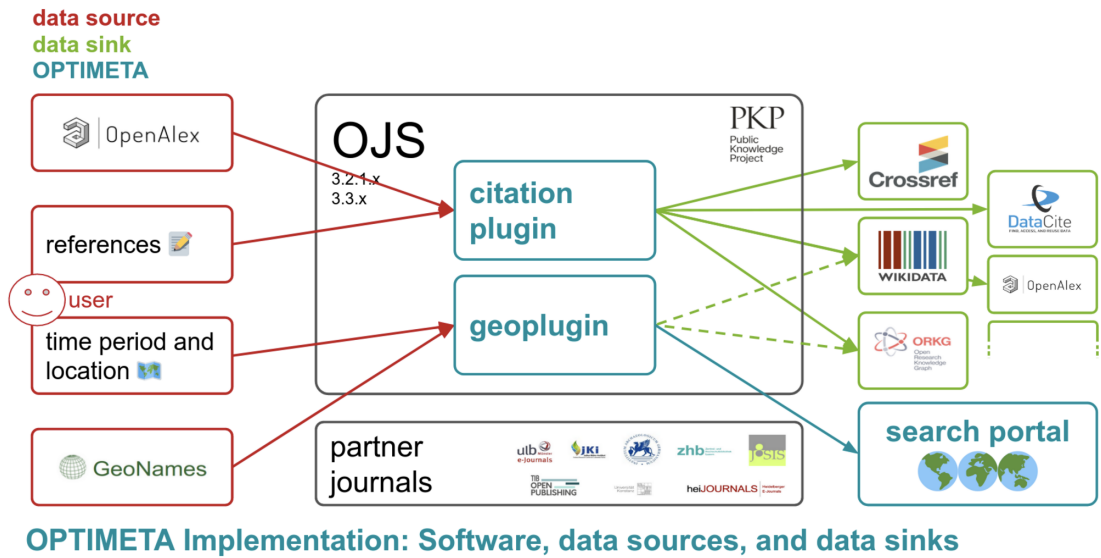

**Fig. 2.** Implementation example of The OPTIMETA Way: OPTIMETA citation plugin and geoplugin for OJS and the connected external data sources, data sinks, and collaboration partners.

### 3.2.2 Citation plugin

The citation plugin aims to gather machine-readable citation metadata during the publication process with the goal of publishing the metadata in open bibliographic data sources. The process is integrated into the existing OJS workflow for submitting publications. When the author begins a submission to the journal, OJS provides a special metadata field for references. We assume that this field is then filled by either the author or the journal editors. The citations are entered into OJS in a raw format, i.e., in a freely modifiable text field (Fig. 3).

> Hauschke C, Nüst D, Cordts A, Lilienthal S (2021) OPTIMETA – Strengthening the Open Access publishing system through open citations and spatiotemporal metadata. Research Ideas and Outcomes 7: e66264. https://doi.org/10.3897/rio.7.e66264

**Fig. 3.** Example of a raw citation.

The citation plugin parses the raw references and extracts the Digital Object Identifiers (DOI) if present. After extracting the DOIs, a look-up algorithm enriches the metadata based on external and open bibliographic data sources in a semantically structured format. Currently, the citation plugin queries the open APIs of Crossref and OpenAlex, which both provide good coverage and promising metadata quality. We have chosen to collect metadata from external sources rather than parsing the citations with parsing tools as we found the results we queried based on the DOI from Crossref and OpenAlex were, in general, more complete and accurate. Developing a custom parsing service or integrating an already existing tool would add unnecessary complexity to the plugin. All of these steps are triggered manually with a single click of a button. Therefore, the additional time needed for **Enriching the scholarly metadata commons**the enrichment of citation metadata comprehends only a few minutes or less, which is neglectable compared to the amount of time consumed for the overall research and publishing process as shown in Fig. 1.

The reliance on DOIs for finding the full reference information is a current limitation of the plugin. The alternative would be to query Crossref with the full reference was not implemented because of the limitations of non-membership access to the API and because the reliable SimpleTextQuery form (https://apps.crossref.org/SimpleTextQuery) is provided for manual use only. We are not aware of a comparable full reference query feature for OpenAlex. In these circumstances, we decided that given the resulting metadata is of higher quality and less manual interaction is required, having a higher degree of automation outweighed the drawback of missing references that do not have a DOI. For new submissions, the plugin will encourage authors to add missing DOIs during the submission process, thus ensuring a reasonably high level of metadata quality and completeness. As OpenAlex harvests from Crossref, having both services as data sources for the plugin may seem superfluous. However, this redundancy avoids being reliant on one specific service into the future and, as OpenAlex harvests other data sources as well, using both sources is likely to provide additional and potentially more complete information.

After these steps have been completed, the enriched results extracted from the external open APIs are presented to the author for review. The review can be done by simply checking whether the results are correct. The various parts of the citations can also be edited manually (Fig. 4). The now semantically structured metadata including title, authors with their corresponding author identifier (ORCID iD), etc. are then stored in the OJS database.

**Fig. 4.** Example of a semantically structured citation which can be edited manually.

After the enrichment process, the citations can be deposited with external services either manually, by clicking the deposit button, or through an automated process managed by the OJS scheduler. The first workflow is currently implemented for OpenCitations. The combined metadata are structured according to Massari and Heibi (2022) and submitted as an issue to a specified GitHub repository of OpenCitations (https://github.com/GaziYucel/open_citations_croci_depot). The issue containing the metadata can then be processed and harvested by OpenCitations.

**Enriching the scholarly metadata commons**

The initial plugin versions focus on using DOIs as the supported publication identifier. In future releases, we are planning to support non-DOI identifiers. Furthermore, being able to import bibliographic metadata via common bibliographic standards like RIS and BibLaTeX would improve the user experience. The current focus of plugin development is to provide support during the publication process for one article. However, journal operators will, naturally, not only want to publish citation information for new articles, but also those from their back catalogue. To address this need, we plan to design a special overview page that will enable articles to be processed in batches. Ultimately, we are also aiming to link into more data sinks as this will enable the widest possible dissemination of citation metadata. For example, we are currently working toward implementation with Wikidata.

### 3.2.3 Spatio-temporal metadata plugin

The OPTIMETA geoplugin enables the collection and display of spatio-temporal metadata for individual research articles in OJS instances. The term "geo" was used in the name as it is more broadly understood and shorter than the more technical "spatio-temporal". Geospatial and geographic data usually include temporal aspects, in that answers to "where in space" questions are always connected to a "when in time" as well. The "geospatial" metadata are more dominant than the temporal metadata in the plugin for different reasons. First, the display of geographical features on a map is more visual and, thus, more interesting than one or several time periods shown as numbers. Second, the novelty and power of geospatial metadata are higher than those of temporal metadata because textual descriptions or classifications of time, such as "in the year 2022" or "during the cretaceous period" are more readily picked up through text searches, whereas spatial relations are harder capture through text descriptions and very difficult to pick up using text searches.

The geoplugin, in particular, extends the *submission workflow*. Due to the simple and intuitive way of entering geo-spatial metadata, the provision of temporal and spatial metadata can be carried out quickly within a few minutes. Authors are asked to provide **temporal metadata** in the form of a simple time period with a start and end date. The date can be entered manually, e.g., by putting in the dates separated by a hyphen into the form field, "2021-01-01 - 2022-02-02", or with an interactive calendar pop-up as shown in Fig. 5. The time period is a simple string and can be used to model temporal uncertainty and allows for imprecision where needed or not known. For example, a history paper may document a hegemony's duration as "753 - 1234", while the observation of a new species may require a precise date such as "2022-08-08 - 2022-08-09".

**Enriching the scholarly metadata commons**

[Screenshot of submission form with Abstract field, List of Contributors, Additional Refinements/Keywords, and Time and location/Temporal Properties section showing a calendar popup for August 2022 and September 2022, with dates 2022-08-01 - 2022-08-10 selected.]

**Fig. 5.** Screenshot of the submission form showing the form field and popup for input of a time period in the lower third of the image.

Next, authors are asked to provide **spatial metadata** using an interactive map as shown in Fig. 6. Authors can add multiple geometries of various types, i.e., points, polylines, rectangles, and polygons. These can be used to adequately represent spatial features related to articles, e.g., places of residence of a historic person, animal tracks, remotely observed areas, or a herd's territory, respectively. An author may choose to quickly add a coarse rectangle providing rather imprecise data or to zoom into the map and enter detailed individual data points, either as time permits or the submission demands. The author is assisted in the creation of this data by two background layers: an open data street map by OpenStreetMap and free-to-use aerial imagery tiles provided by Esri. In this way, both natural features and human-built structures can provide orientation. While creating the geometries forming the detailed spatial metadata, the geoplugin constantly queries the Geonames gazetteer service (https://www.geonames.org/) using coordinates from the geometries to automatically derive the bounding rectangle or bounding box of the smallest encompassing

**Enriching the scholarly metadata commons**

administrative area. The administrative unit is given in a form field below the map. The data for administrative units varies widely for different countries around the world and is, arguably, most useful for countries in the global north.

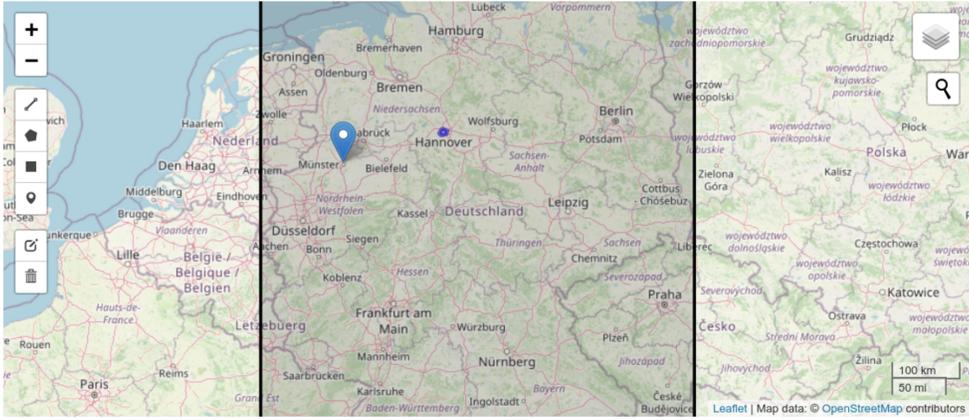

**Fig. 6.** Screenshot of the submission form showing the interactive map for collecting spatial metadata, in this case a point geometry in City of Münster and a polygon around the city of Hanover in blue; these geometries are enclosed in the administrative unit "Earth, World, Germany" whose bounding rectangle is shown on the map in black.

The information collected during the submission can then be reviewed during the editorial process. Ultimately, the data is stored as plain text in the OJS database in GeoJSON (https://geojson.org/) format. A single `FeatureCollection` includes the geometries and a short provenance statement indicating who created the data or where it was derived from. We chose GeoJSON for this purpose despite the limitations resulting from its inability to handle different coordinate reference systems because of its wide usage and simplicity. For the purpose of discovering research articles on a global scale, the accuracy of several metres of the coordinate reference system "World Geodetic System 1984" (WGS 84) is entirely sufficient, especially considering that most geometries are manually created on an interactive map.

**Enriching the scholarly metadata commons**

**Fig. 7.** Screenshot of the article landing page with spatio-temporal metadata; this article has multiple polygons covering parts of Brazil with a matching textual description below the map, but there is no time period; the column on the right contains the download button for metadata in GeoJSON format.

**Enriching the scholarly metadata commons**

**Fig. 8.** Screenshot of an article landing page with publication information and spatio-temporal metadata; above the map the time period of interest is given, the geometries describing this article's content are several polylines representing travel routes.

The spatio-temporal metadata is then published together with the article on various pages: the article landing page (see Fig. 7 and 8), the landing page for the issue (see Fig. 9), and on a separate page for the journal itself. The article landing page also contains a download button providing easy access to the spatio-temporal metadata in GeoJSON format. The journal landing page features synchronised highlighting as shown in Fig. 9 if the user holds the cursor over a geometric feature on the map. Both the feature on the map and the corresponding article in the list above is highlighted in red. Clicking on the issue or journal map opens a small popup window (not shown) containing the publication metadata (author, title, etc.) and a link to the article landing page. Below each map, there is a clear

**Enriching the scholarly metadata commons**

statement about the licence of the spatio-temporal data, to which authors will need to agree to while creating the metadata. The licence is fixed to a public domain licence, CC-0, to ensure the broadest possible usage.

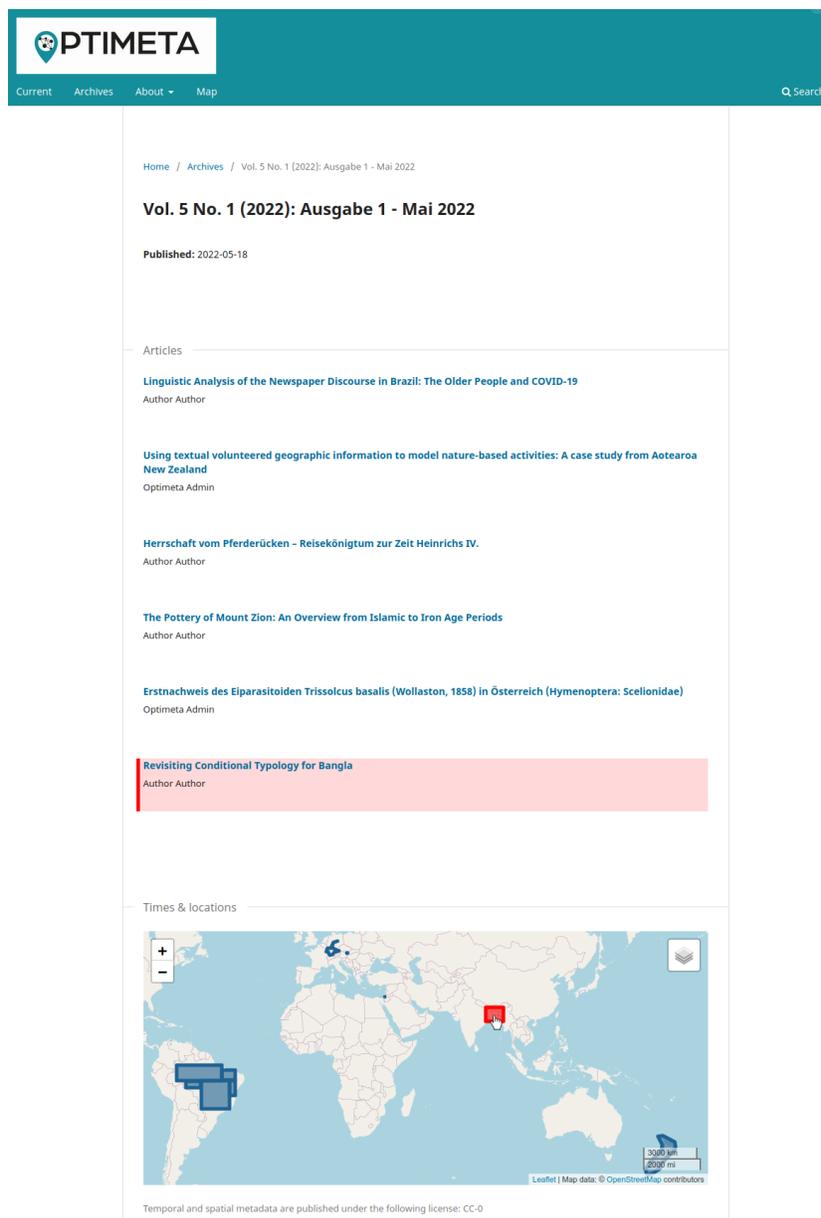

**Fig. 9.** Screenshot of the issue view in the public demo journal, see https://service.tib.eu/optimeta/index.php/optimeta/issue/view/1. The standard OJS theme is extended with a "Times & locations" section below the list of articles of the issue. The mouse cursor over the spatial feature on the map at the bottom triggers a highlighting of the geometry and corresponding article in the list above in red.

The pages shown above, all target human users, but the spatio-temporal metadata are also included in the HTML website of the article landing page in machine-readable form. These metadata fields

**Enriching the scholarly metadata commons**

enable scraping and harvesting through other services. Fig. 10 shows selected values as displayed in the HTML header of a test article, each defined by a name and, if available, a well-defined scheme. Alongside other publication metadata, the spatial metadata is included in several forms and schemas including the Dublin Core fields `DC.SpatialCoverage` and `DC.Coverage`. The former is included as a textual encoding of the full GeoJSON record (line 10 in Fig. 10), the latter (line 20) as a textual representation of the administrative units starting with the largest units and working down to the more generic field, `geo.placename`, which contains the smallest available administrative unit. In the example provided, this is a country name, but it can also be more specific, for example, the name of a town. Finally, the bounding rectangle of the smallest administrative unit is given in the fields `ISO 19139` in an XML-encoding of the geographic bounding box according to the ISO 19139 standard and `DC.box` using a simple list of the coordinates of the four cardinal directions limiting the rectangle separated by semicolons. The temporal metadata is stored in the field `DC.temporal` and `DC.PeriodOfTime`, both using textual representations of a time period as defined by ISO8601.

The initial development phase focused on the collection of metadata during submission and the display of spatial metadata. Later phases will focus on the development of a more sophisticated and interactive display of the temporal metadata, specifically, putting the time period(s) of papers on a common timeline for specific issues and whole journals, adding support for multiple time periods, increasing the range of historic date formats supported (BC, time frames of millions of years, etc.), building in a function for entering coordinates directly, support for personalised reference datasets (related to a journal's themes/topics, e.g., biospheres, habitats) for use in spatial metadata creation, and deriving spatio-temporal metadata semi-automatically, e.g., by retrieving information from data deposits or examining data files in supplementary materials.

```html
1  <!DOCTYPE html>
2  <html xml:lang="en-US" lang="en-US"><head>
3  <meta http-equiv="content-type" content="text/html; charset=UTF-8">
4  <meta charset="utf-8">
5  <meta name="viewport" content="width=device-width, initial-scale=1.0">
6  <title>Test 3: Three | Journal of Optimal Geolocations</title>
7
8  <meta name="generator" content="Open Journal Systems 3.3.0.11">
9  <meta name="DC.temporal" scheme="ISO8601" content="2022-06-27/2022-06-30">
10 <meta name="DC.SpatialCoverage" scheme="GeoJSON" content="{"type":"FeatureCollection","features&qu
11 <meta name="geo.placename" content="Italian Republic">
12 <meta name="DC.box" content="name=Italian Republic; northlimit=47.091783741544; southlimit=35.49285259236; westlimit=6.6266
13 <meta name="ISO 19139" content="<gmd:EX_GeographicBoundingBox><gmd:westBoundLongitude><gco:Decimal>6.6266
14 <meta name="DC.PeriodOfTime" scheme="ISO8601" content="2022-06-27/2022-06-30">
15 <meta name="gs_meta_revision" content="1.1">
16 <meta name="citation_journal_title" content="Journal of Optimal Geolocations">
17 <meta name="citation_author" content="C Contributor">
18 <meta name="citation_title" content="Test 3: Three">
19 <link rel="schema.DC" href="http://purl.org/dc/elements/1.1/">
20 <meta name="DC.Coverage" xml:lang="en" content="Earth, Europe, Italian Republic">
21 <meta name="DC.Creator.PersonalName" content="C Contributor">
22 <meta name="DC.Title" content="Test 3: Three">
23 <meta name="DC.Type" content="Text.Serial.Journal">
24 <meta name="DC.Type.articleType" content="Articles">
```

**Fig. 10.** Screenshot of the source code of the article landing page showing selected HTML meta attributes given in the HTML header, including different representations of spatial and temporal metadata.

## 3.3 Enriching the scholarly metadata commons

We conceptualise the scholarly metadata commons as a special subset of the knowledge commons (Hess and Ostrom 2006; Mansell 2013), in which an openly licenced and, thus, collectively owned aggregation of scholarly metadata is governed and shared among the community of interested scholarly and related stakeholders. This commons has various manifestations that present data in a user-friendly interface, in the form of websites or APIs, and enable both the contribution and extraction of data. Wikidata is a widely known example of such an interface.

We make use of and contribute to the Scholarly Metadata Commons through both plugins:

1. Citations plugin: With this plugin, we expand the open data pool for research information by providing enriched and user-verified metadata, collected and distributed at the time of publication, through open APIs. We also incorporate sources not currently included in the standard scientometric data sources because of their language or because they are not supported by big publishing houses.
2. Spatio-temporal metadata plugin: Through this plugin, we enable new use cases such as location-based assessments of research activities and location-based research discovery, based on, for example, (1) questions about the geographical area being studied and (2) new transdisciplinary connections between research outputs based on time periods and areas of interest beyond commonly used keywords and full-text search.

## 4 Discussion

The OPTIMETA Way described above provides three important contributions, which we implement here as exemplary with the presented plugins. The first contribution is the **enlargement of the Scholarly Metadata Commons** with metadata generated during the publication process. The built-in mechanisms for looking up existing metadata and the following import of persistent identifiers, such as ORCID iDs, enable the creation of strongly linked research information and its subsequent exportation into existing data sinks. While non-English publications play an important role in the academic world (Kulczycki et al. 2020; Liu 2017; Nazarovets and Mryglod 2021), their metadata are not currently equally represented in the major citation databases (Tennant 2020; Vera-Baceta et al. 2019).

The second contribution is, to allow for many more **Open Access journals in citation databases** and other services built upon the scholarly metadata commons. Currently, OJS is being used by more than 25,000 journals from 155 countries (the majority being from the Global South) publishing in 56 languages (Khanna and Willinsky 2022). Using our plugin will lower the barrier for independent journals to contribute to open bibliographic metadata considerably, albeit currently only if the journals use OJS. While this is a large step towards a solution, we cannot yet eliminate the problem entirely. Therefore, we hope that The OPTIMETA Way will be implemented in other publication platforms in the future.

**Enriching the scholarly metadata commons**

The third contribution is the **expansion of the scholarly metadata commons through the inclusion of spatio-temporal metadata**, which facilitates the use of open metadata for new use cases. As Niers and Nüst (2020) explain, spatio-temporal metadata can be used to detect biases in the geographic coverage of research, for example, when research in a given field focuses heavily on one region overlooking other areas that may be no less interesting in the process. Spatio-temporal metadata can also help recognise connections between research works and improve the understanding of geographical and time-based relations within an area of study. Furthermore, visualisations, especially in the form of maps, can support the transfer of research content and the need for research as a whole. To date, the availability of spatio-temporal metadata has remained low and with the release of geoplugin, we will contribute a new component to the ecosystem of open scholarly publishing. Furthermore, the geoplugin will enable new use cases in location-data-based assessment of research activities: (1) answering questions about the area that has been investigated, e.g., to demonstrate a specific coverage or distribution of research locations and (2) detecting potentially valuable transdisciplinary connections between research outputs based on time periods and areas of interest that go beyond commonly used keywords and full-text search, e.g., connecting historical works on social questions in central Europe with current research on health. In the future this metadata can be used to build platforms for timely notifications about publications based on user-defined spatio-temporal interests, i.e., so that users or systems can be notified of new publications that cover an area of particular interest to an assessment scheme. Intentionally imprecise coordinates can be used to preserve the privacy of human subjects or hide protected entities.

The **integrative power of spatial relationships** between research articles has already been acknowledged by others. However, none of the existing solutions follow The OPTIMETA Way and are, therefore, too complex, not integrated into the publishing workflow, or do not contribute to the open scholarly metadata commons. For example, the JournalMap (https://www.journalmap.org/; Karl et al. 2013) shows research paper locations and publication metadata (title, abstract, etc.) for map-based discovery. However, JournalMap is limited to point geometries and while there is an API and some collaboration with publishers (https://www.journalmap.org/publishers; https://web.archive.org/web/20161016000907/https://newsroom.taylorandfrancisgroup.com/news/press-release/taylor-francis-journal-map-partnership#.WALFJmF_o88), the data is not fully open. The announcement that a data download option is "coming soon" has been on the website since its inception (see https://web.archive.org/web/20130615020154/https://www.journalmap.org/downloads) and the licence is defined as Creative Commons Attribution Share-Alike (CC-BY-SA, https://www.journalmap.org/developer/documentation/1-0), but the terms of use then limit the licence terms considerably and prohibit commercial use of the data (https://www.journalmap.org/terms-of-use). The website does offer some advanced filtering options, including additional thematic filtering options. However, the commercial options advertised on the website work against our understanding of knowledge advancement. Second, Kmoch et al. (2018) analysed articles from geoscientific journals to automatically derive spatial metadata from the unstructured information in articles' bibliographic metadata. The extracted data was then published in a public geospatial catalogue service. However, this approach required considerable technological knowledge and lacked human quality checks, as not all data was checked by the most suitable experts. Therefore, despite being a valid approach for dealing with the fact that spatio-temporal metadata was not collected in the past, it is neither a complete solution nor in line with The OPTIMETA Way. Garzón and Nüst (2021b) took a similar approach with the tool geoextent, which they used to create a discovery index based on geospatial metadata for generic research data repositories. They used a brute-force approach to retrieve spatial extents from as many geospatial file formats as possible. This could be an intermediary approach to enrich article metadata if the articles properly cite the data used, though human verification would likely be needed as datasets may be cited for many different reasons. An implementation of the search portal (cf. Fig. 2) that collects spatio-temporal metadata from multiple journals, the OPTIMAP, is currently under development (see https://optimap.science/ and https://github.com/ifgi/optimetaPortal).

The **advantages offered by the availability of open citation information** are undeniable. Peroni and Shotton (2020) provide an extensive list of beneficiary stakeholders: researchers who do not belong to "the elite club of research universities that can afford subscription access to the commercial citation indexes WoS and Scopus", bibliometricians, who want to provide research data on their research, librarians, funders, research managers, and much more. A key value of the citation plugin is that it allows a large set of Open Access journals to share their authors' publications in the open research commons, regardless of language or subject area and whether it is a publisher-led or independent scholar-led journal. The often shamefully overlooked long-tail of academic research will thus become visible and be given the opportunity to be properly integrated, especially the often overlooked non-English literature (Lazarev and Nazarovets 2018).

With respect to **science communication and assessment**, Krüger (2020) describes how the social distrust of science is to be countered by a performativity-measuring quantification of research output and associated metadata and indicators. She argues, convincingly, that bibliometric infrastructures and applications have their own ideas about how research can be understood through their use. By expanding the scope and range of what we have in terms of metadata, The OPTIMETA Way cannot fully prevent this, but it can address the extent of the problem by limiting the distorted perception of what can be observed, measured, assessed, and considered knowledge through its digital representation in metadata. In this way, the improved availability of spatio-temporal and citation metadata means that research assessment can be carried out more quickly, easily, transparently, and responsibly. **Desiderata in terms of metadata** could be machine-actionable descriptions of research problems, methods, connections to external entities like funding IDs or funders, identifiers for physical samples, or identifiers for instruments.

We estimate the risk of **unethical use** of the plugins is low and do not see particular potential for the misuse of spatio-temporal data. Even in the worst-case scenario, while intentionally defective metadata reduces discoverability, it does not impact other means of identifying publications. It is true that, in relation to citation metadata, the plugins do lower the barrier to depositing falsified citation information when irresponsible research assessment methods such as citation counts are of interest to a malevolent party. However, citation metadata can only be misused if both author and the handling editor, who is encouraged to check the citation information before publication, have malicious intent. Furthermore, the deposition in public databases does not happen anonymously and, once identified, any misuse can be rolled back and accounts can be blocked from uploading further data.

**Enriching the scholarly metadata commons**

In the **future**, the implementations of The OPTIMETA Way presented here could be extended with new features and improved usability based on the experiences reported by journals from different disciplines. Regarding technology, we imagine more sophisticated methods, such as machine learning approaches or the extraction of information from PDFs and data files, could be integrated into the OPTIMETA plugins to increase the usability and extent of the metadata. For example, acknowledgements such as funding bodies and grant IDs, author contributions (e.g., based on an acknowledgement section using CRediT statements), or subject classifications, could be collected in a similar, semi-automatic way and publicly deposited according to The OPTIMETA Way. However, to protect the quality of the data, validation by a human expert should not be omitted. The scope of the enhancement with metadata can be widened to include preprints, monographs, and edited collections, although semantically meaningful attributes to metadata fields will be required to distinguish non-reviewed research outputs from reviewed ones. To support preprints and books, the citation plugin and geoplugin can be ported to PKP's preprint platform in addition to other book publishing platforms.

The shift in granularity and speed that can be expected due to more open and also more pressing research as societal **challenges** are tackled in the future will require even more timely, validated research metadata for effective communication. Research is increasingly being published in stages (e.g., Octopus, https://www.octopus.ac/about) and as individual building blocks (e.g., idea/text/interpretation, code/software, data), rather than as a polished textual artefact years after the issue might already be resolved. Researchers will more regularly share their results accessibly on free infrastructures and peer review practices will adapt, e.g., with overlay journals (Brown 2010; Rousi and Laakso 2022). However, these technical challenges are small when compared to the organisational challenges of ensuring the long-term maintenance of the plugins we have developed. While the current funding facilitated the development of stable plugins and provided for them to be sent to select collaboration partners for evaluation, and while more and more programs funding core research software are being founded (e.g., https://chanzuckerberg.com/eoss/), we are still facing a chicken-and-egg problem. For broad uptake, journals require a commitment to long-term software maintenance, while funding to maintain the plugins and improve them is acquired more easily when broad usage can be demonstrated.

The cultural shift towards Open Access and FAIR research information housed in open infrastructures (Hauschke et al. 2021a; Hendricks et al. 2021) will happen at different speeds in different countries and disciplines and result in the coexistence of a variety of platforms. This is an advantage over today's centralised system and the power large publishers have over it, but it is also a challenge as these services will have to be able to connect and exchange metadata. Therefore, it is our hope that The OPTIMETA Way will be transferred to other elements within the academic open infrastructure, so that targeted, novel, and even small scale metadata attributes can be collected from the most knowledgeable party, with minimal impact on existing workflows, and shared broadly, openly, and quickly for the advancement of knowledge.

## 5 Conflict of Interest

The authors declare that the research was conducted in the absence of any commercial or financial relationships that could be construed as a potential conflict of interest.

## 6 Author Contributions

The authors contributed equally to this work. Development of the described plugins was conducted by Daniel Nüst (spatio-temporal metadata plugin) and Gazi Yücel (citation plugin).

## 7 Funding

The authors acknowledge the financial support by the Federal Ministry of Education and Research of Germany (BMBF) in the framework of OPTIMETA (grant numbers 16TOA028A and 16TOA028B).

## 8 Acknowledgements

The authors would like to thank our project partners for the continuous discussions on how to improve the OPTIMETA Way. In particular, special thanks go to Open Citations and PKP for their supportive engagement in technical discussions. We thank Tom Niers for developing the first prototype of the spatio-temporal metadata plugin, and Svantje Lilienthal for contributions to the early conceptual discussions on the citations plugin. We thank Julie Davies from the academic editing service of the University of Münster for her support in revising the manuscript.

# Enriching the scholarly metadata commons

# Enriching the scholarly metadata commons